\documentclass[aps,twocolumn,showpacs,prl,preprintnumbers,amsmath,amssymb]{revtex4-1}

\usepackage{graphicx}
\usepackage{dcolumn}
\usepackage{bm}
\usepackage{amsmath,amssymb}

\begin{document}

\title{Heterogeneous structure of granular aggregates with capillary interactions}

\author{Michael Berhanu}

\author{Arshad Kudrolli}%

\affiliation{%
Department of Physics, Clark University, Worcester, Massachusetts 01610
}

\date{\today}

\begin{abstract}
We investigate the spatial structure of cohesive granular matter with spheres floating at an air-liquid interface that form disordered close packings with
pores in between. The interface is slowly lowered in a conical container to uniformly compress and study the system as a function of area fraction $\phi$. We find that the free area distributions associated with Voronoi cells show significant exponential tails indicating greater heterogeneity compared with random distributions at low $\phi$ with a crossover towards a $\Gamma$-distribution as $\phi$ is increased. Further, we find significant short range order as measured by the radial correlation function and the orientational order parameter even  at low and intermediate $\phi$, which is absent when particles interact only sterically. 
\end{abstract}


\maketitle

The effect of cohesive interactions on the structure of granular matter is an important question which has bearing on its strength and stability. Further, aggregation of cohesive particles is a broad physical phenomenon occurring among others in colloidal systems, dust, and soot due to electrostatic interactions, inter-galactic dust under gravitation, and in powders due to humidity. Recent numerical studies have shown that the connectivity and rigidity transition for cohesive frictionless  particles occur at distinct and lower  concentration compared with non-cohesive particles~\cite{Lois}, and that cohesion and friction leads to more porous structures~\cite{Head,Gilabert}. While there have been some experiments which have examined the effect of cohesion on packing of wet granular matter at the particle level in the dense regime~\cite{Xu}, there have been no experiments which have examined structure over a broad range of packing fraction. Complex structures have been observed with magnetic beads, but the interactions there are dipolar~\cite{Blair2}. 
\begin{figure}
\includegraphics[width=8cm]{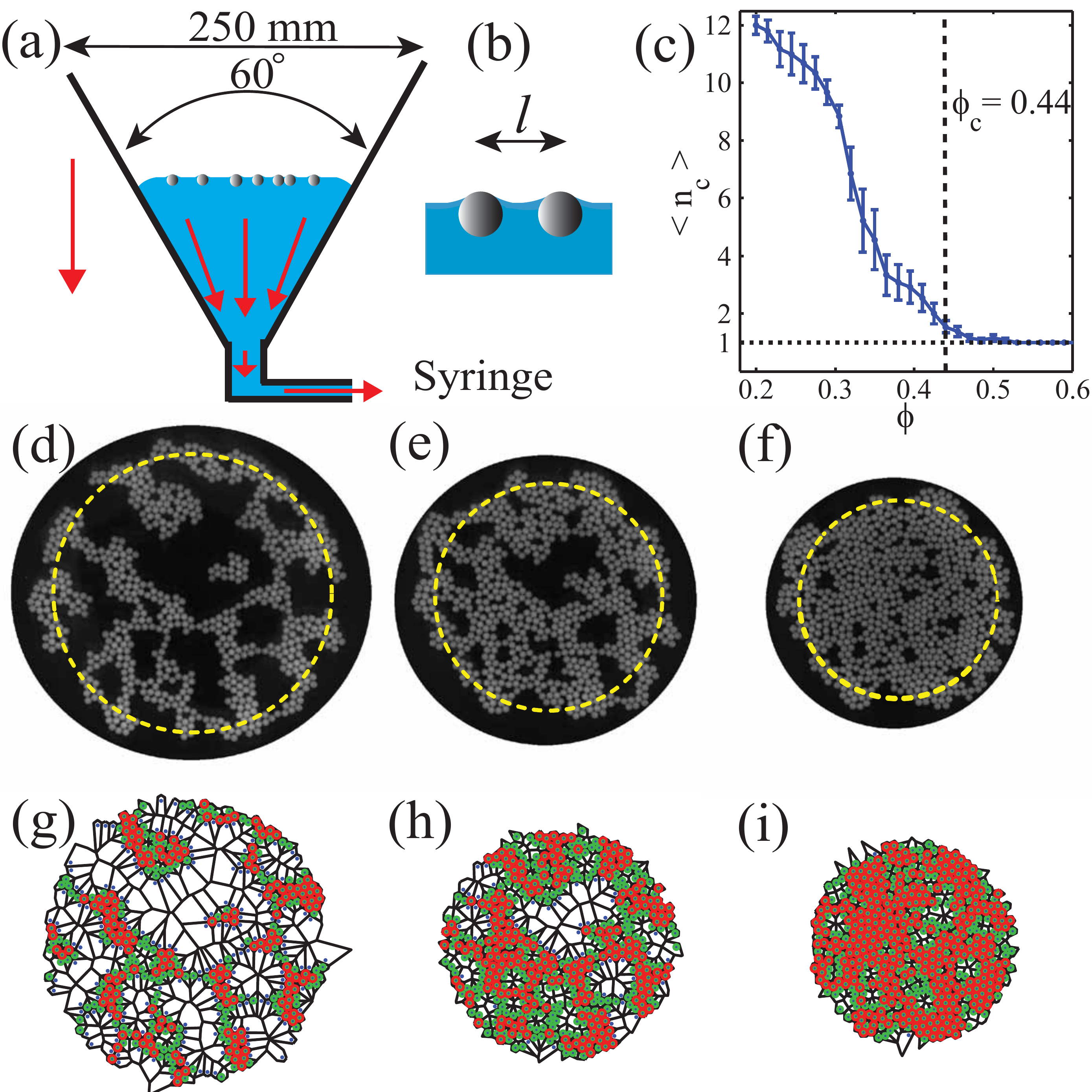}
\caption{\label{fig1} (a) Schematic diagram of the experimental apparatus. The liquid level inside the funnel is decreased to increase the area fraction $\phi$ of the floating spheres. 
(b) The meniscus around floating spheres leads to attraction. 
(c) Mean number of individual aggregates versus $\phi$. Only one aggregate remains on average above $\phi_c=0.44$. Sample images of floating spheres ($N=688$) observed at $\phi=0.33$ (d), $\phi=0.50$ (e), and  $\phi=0.65$ (f), and corresponding Voronoi cells inside the dashed circle (g), (h), and (i). Cells corresponding to tightly packed regions using criteria $d_v/d < 0.1$ and $0.1 < d_v/ d < 0.5$ are identified with dark grey/red and grey/green markers, where $d_v$ is the distance of the particle from the center of the polygonal Voronoi cell. 
}
\end{figure}

Here, we use capillary attraction between spheres floating at the air-liquid interface~\cite{Chan,Cheerios} as a model granular system with isotropic cohesive interactions. Because of the exponential nature of attraction between floating particles with distance, the interaction is essentially short range for millimeter sized particles. Further, a minimum force is required to overcome the rolling or sliding friction because of the cohesive normal force acting on particles in contact. Thermal excitation is insignificant at these scales, and therefore floating spheres can be considered as a model system to understand the effect of particle level interactions on the heterogeneity of cohesive granular aggregates without  the complexity introduced by gravitational gradients. Recently, such a system was used to study the granular character of particle rafts by compressing them at the boundaries using a Langmuir trough~\cite{cicuta09}, but neither the area fraction nor the particle level heterogeneity was examined. 

A novelty of our apparatus is that we can uniformly change the particle concentration at the air-liquid interface by varying the height of a highly viscous liquid inside a conical container. Thus, any two material points on the interface approach each other proportional to distance of separation as the liquid level is lowered, rather like two points on the surface of an inflated balloon from which air is released. Using this trick, we compress and vary density of the entire system homogeneously. By imaging and finding the positions of the particles, we calculate statistical measures including Voronoi cell area distribution, and the radial and angular correlations to characterize the heterogeneity and understand the interplay of cohesion and steric interactions. 

A schematic of the experimental apparatus is shown in Fig.~\ref{fig1}(a) and consists of a conical glass funnel filled with a $90-10$\,\% glycerol-water liquid mixture by mass.  The liquid has a density  $\rho_l = 1.235 \times 10^{3}$\,kg-m$^{-3}$, dynamic viscosity $\eta=0.26$\,Pa-s, and surface tension $\gamma=6.5 \times 10^{-2}$\,N-m$^{-1}$. The particles used in our experiments consist of white polyethylene spheres of diameter $d=3.175$\,mm and density $\rho_p =0.95 \times 10^{3}$\,kg-m$^{-3}$. The funnel is treated chemically to create a hydrophobic boundary conditions which repels the floating particles to about 2\,cm from the boundaries. Images of the floating spheres are recorded with a $1392 \times 1040$\,pixel CCD camera, and centers located to within $0.1\,d$ using standard particle tracking software. 

The experiments are initialized by first placing $N$ spheres at random locations, which are then observed to rearrange over a time scale of a few seconds and form small aggregates that remain in place for hours~\cite{allain88}. This can be explained by the fact that the force between  spheres separated by a distance $l$ (see Fig.~\ref{fig1}(b) for small deformations is given by~\cite{Chan,Cheerios}  $F(l) \propto - (l/L_c) ^{-\frac{1}{2}} \, \exp(- l/L_c)$ for $l/L_c \gg 1$, where the capillary length $L_c = \sqrt{\gamma/\rho_l\,g } \sim 2.3$\,mm $< d$. By balancing capillary and drag forces one finds that time over which spheres approach each other increases roughly exponentially with distance. 
Next, after a waiting time of 600\,s, the liquid level is slowly lowered at a constant rate using a syringe as indicated in Fig.~\ref{fig1}(a) over a time  of 300\,s~\footnote{Movies of typical experiments and further information on uniformity of compression and analysis of shape factor of Voronoi cells can be found in the supplementary material}. Over 600\,s the range of attraction is about $4d$, and during the compression, the motion for isolated spheres due to capillary forces occurs only over roughly $0.2d$. Because of the symmetry of the conical funnel, the radial velocity of a liquid element at the air-liquid interface away from the boundaries is approximately given by  $\vec{v}(r) \sim - C_0\, \overrightarrow{r}$, where $C_0$ is measured to be $3 \times 10^{-3}$\,s$^{-1}$ using a tracer technique and $\overrightarrow{r}$ is the radial vector to the center axis of the funnel. Such a flow results in isotropic homogeneous compression of particles floating on the surface, because the strain is independent of position. Further, inertial effects are negligible because the maximum Reynolds number calculated using the particle size is approximately $5 \times 10^{-3}$. 

Examples of aggregates observed are shown in Fig.~\ref{fig1}(d-e-f) at various stages of compression. The corresponding area fraction $\phi$ is obtained as the fraction of area occupied by the spheres inside a circle (yellow dashed line) whose center is the mean position of spheres and radius is chosen as $\sqrt{3/2}$ times the radius of gyration in order to avoid any boundary effects. Voronoi tessellation is computed using Qhull programs in MATLAB in the bulk indicated by the dashed circle indicated in Fig.~\ref{fig1}(d-e-f). No particular variation in size with distance from center can be noted supporting our claim that the compression is uniform~[10].

The aggregates initially appear to phase separate into regions with high density of particles with voids or pores in between. As the system is compressed, the number of aggregates decreases as shown in Fig.~\ref{fig1}(c) till a threshold is reached where only one connected aggregate remains. This connectivity threshold is found to occur at $\phi_c = 0.44 \pm 0.02$ by averaging over several experiments. Upon further compression, void spaces decrease in size, until a critical area fraction $\phi_j = 0.715 \pm 0.006$  is reached where the aggregates appear to buckle, and the system is no longer two-dimensional. This buckling transition has been studied previously~\cite{cicuta09}, and may be considered as the point where jamming occurs for this system because the system can no longer accommodate in plane deformation upon further compression. With this assumption our data provide evidence that the connectivity transition occurs at a distinct and lower $\phi_c$ compared with $\phi_j$~\cite{Lois}, while the values themselves are not universal and depend on the strength of interaction. This is in contrast with a non-cohesive system where these points coincide~\cite{Lois}. Finally, it is noteworthy that $\phi_j$ is significantly lower than observed with frictional and cohesionless disks~\cite{Majmudar}, where $\phi_j \sim 0.84$.

\begin{figure}
\includegraphics[width=8.cm]{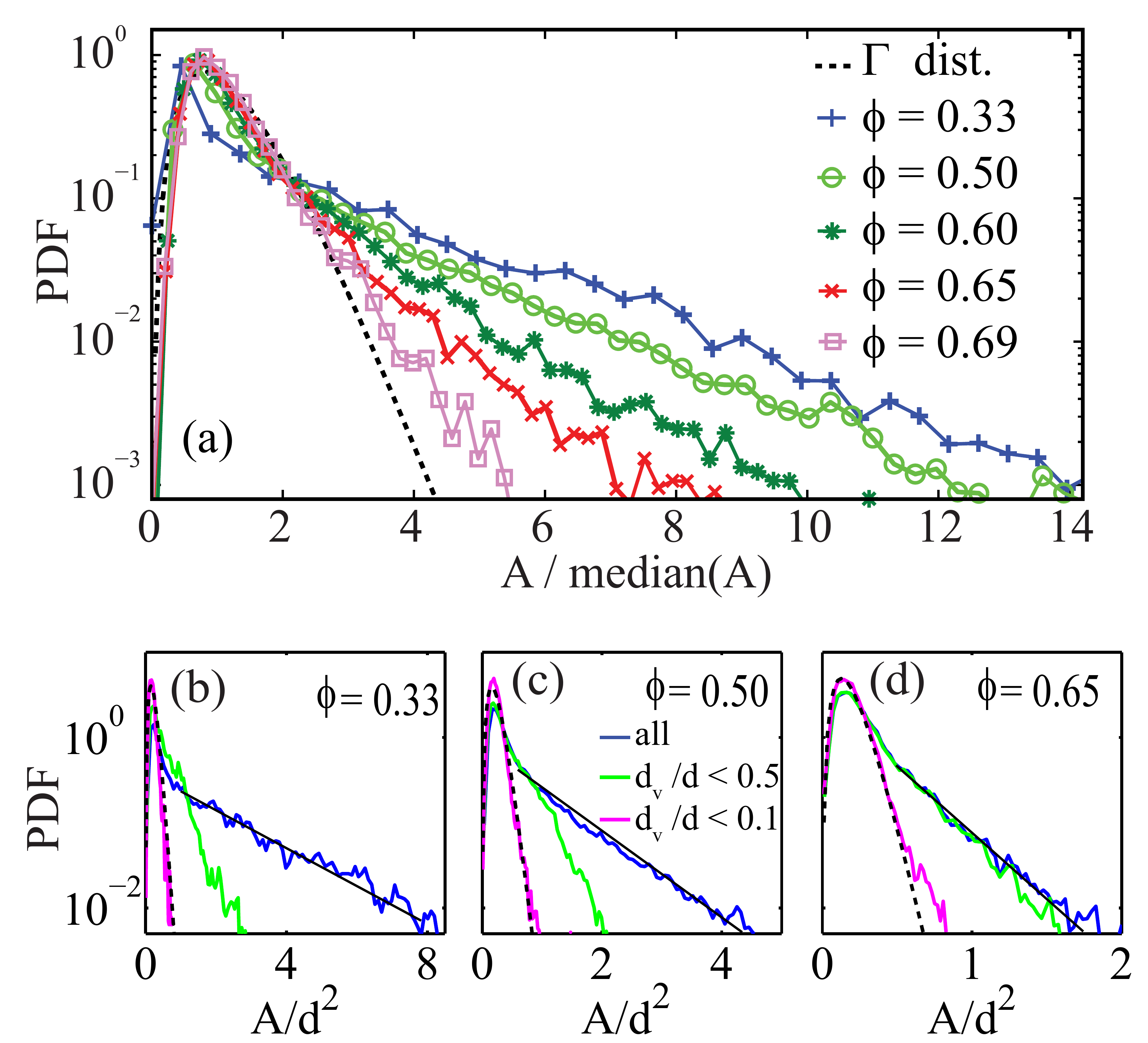}
\caption{\label{fig2} Probability distribution function (PDF) of the free area $A$ normalized by the median value of $A$ is observed to deviate from a $\Gamma$ distribution ($\nu = 3.6$) postulated for random distribution of points. The deviation is strongest at low $\phi$ with significant probability of finding large cells. (b-c-d) The tail of the free area PDF with all the cells can be fitted by an exponential function with a decay parameter $\alpha=0.484$ (b), $\alpha=1.11$ (c), and $\alpha=2.82$ (d). When Voronoi cells are selected with  $d_v/d < 0.5$ the exponential tail decreases, and the PDF approaches a $\Gamma$ distribution ($\nu = 3.6$) for $d_v/ d < 0.1$.}
\end{figure}

In order to examine the heterogeneity of the packing, we analyze the size distribution of the Voronoi cell~\cite{Schenker}. In case of random uncorrelated points, the distribution of the area of the Voronoi cell $A_v$ has been argued to be a $\Gamma$ distribution~\cite{Weaire}:
$$P(A_v) \sim A_v^{\nu-1} \, \mathrm{exp} \left( -\nu \dfrac{A_v}{<A_v>} \right),$$ 
where, $\langle A_v \rangle$ is the mean area and the shape parameter $\nu \sim 3.6$ from geometric arguments. Now for hard disks, a free area $A$ is introduced, which is given by $A = A_v - \dfrac{\sqrt{3}}{2} d^2$, the minimum possible cell area corresponding to a hexagonal packing. Numerical simulations~\cite{Kumar} with frictionless hard disks have shown that the free area distributions are consistent with a $\Gamma$ distribution, and experiments with steel cylinders~\cite{Lechenault} have shown $\nu$ between $3$ and $4$ depending on the vibration and the roughness of cylinder. From Fig.~\ref{fig1}(d-f-h) large variation in the sizes of the cells are apparent at low $\phi$, but less so as the $\phi$ is increased reflecting the changes in the void spaces between particles. To quantify this distribution, we obtain the probability distribution function (PDF) of $A$ and plot it in Fig.~\ref{fig2}(a) as a function of area normalized by the median.  We find significant changes in the shape as a function of $\phi$ with systematically greater deviations from $\Gamma$ distribution at lower $\phi$. 

To understand these deviations, we note that the particle associated with a Voronoi cell appears more asymmetrically placed if the particle is located at the edge of a large pore. Using this observation, we define a criterion based on the distance of the particle from the center of the polygonal cell area $d_v$ to divide the particles and their associated Voronoi areas in to subsets. In Fig.~\ref{fig1} (g-h-i), we have indicated a subset of particles with a grey/green using a criteria $0.1 < d_v/d < 0.5$, and a subset using a criteria  $d_v/d < 0.1$ with red/dark grey. We find that all areas with substantial overlap with large void spaces are eliminated in this last subset. We then plot PDF of areas with $d_v/d < 0.5$ and subset with $d_v/d < 0.1$ along with PDF of all the areas in Fig.~\ref{fig2}(b-c-d) for the three representative $\phi$. We find that the PDF for $d_v/d < 0.1$ approach a $\Gamma$ distribution with $\nu=3.6$, whereas PDF corresponding to all the cells show strong exponential tails. Thus, we are able to link the nature of the distribution of free area for all particles to the  presence of heterogeneity in the system due to the presence of the voids. Additional complementary analysis  shows that cells become more regular shaped with compression and is discussed in the supplementary documentation~[10].

In order to characterize the void spaces or pores directly, we investigated the 
pore size distribution using the distance of a random point in a pore to the nearest point on the pore-solid interface~\cite{Torquato}. As shown in the supplementary documentation~[10], we find a wide distribution of pores size whose average size decreases continuously during the compression. This trend reflects the evolution of the exponential tail in free area distributions $\sim \mathrm{exp}(-\alpha A)$, where the parameter $\alpha$ is related to the average pore size.

\begin{figure}
\includegraphics[width=8.2cm]{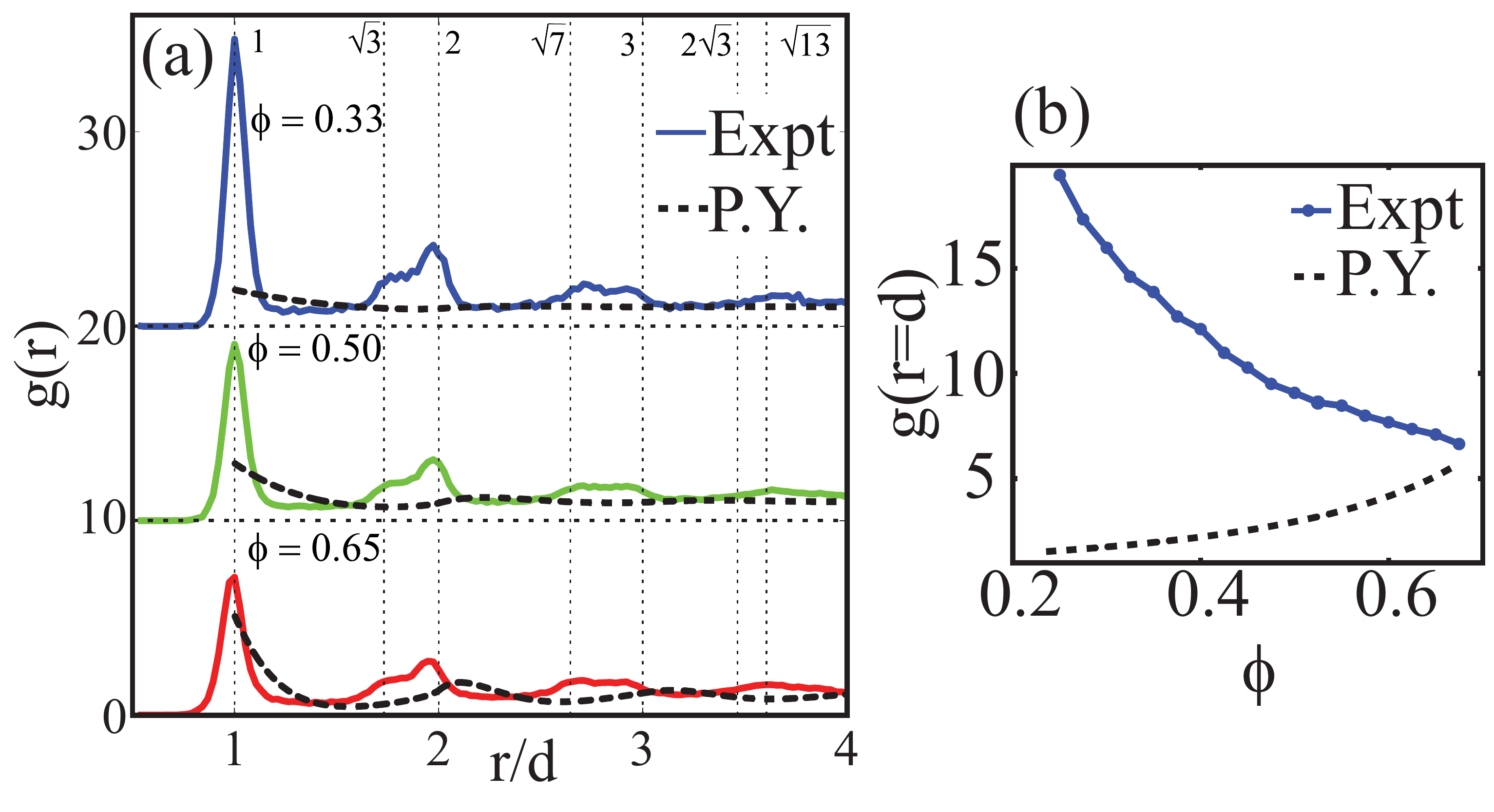}

\caption{\label{fig3} (a) The radial correlation function $g(r)$ for $\phi=0.33$, $\phi=0.50$, and $\phi=0.65$ (Top to bottom). Experimental results are compared with theoretical calculation for hard disks given by the Percus-Yevick equation. (b) $g(r=d)$ (amplitude of the first peak of this function) versus $\phi$ for experiments with attractive particles and for solution of Percus-Yevick equation.}
\end{figure}

To examine the short scale structure in the aggregates, we evaluate the radial correlation function $g(r)$ which measures the particle density-density correlation at distance $r$ in Fig.~\ref{fig3}(a) for various $\phi$. For hard particles with diameter $d$, $g(r)$ = 0 for $r < d$, and $g(r) \rightarrow 1$ for large $r$. Peaks are clearly observed near $d$, $\sqrt{3}d$, $2d$, $\sqrt{7}d$, $3d$, $2\sqrt{3}d$ and $\sqrt{13}d$ indicating a hexagonal packing. Further it may be noted that the locations of the peaks do not vary significantly as $\phi$ is increased. These observations appear to be consistent with the fact that floating particles are pulled towards each other if they are close by until they come in contact. As $\phi$ is increased, the relative height of the secondary peaks increase consistent with an increase in size of the clusters. 

Now, it is well known that $g(r)$ for non-cohesive hard-disks show peaks because of steric interactions as the density is increased. Therefore, we also plot in Fig.~\ref{fig3}(a) $g_{P.Y.}(r)$ calculated for hard disks arranged randomly using the Percus-Yevick equation~\cite{Percus2}. The peak at $r = d$  for cohesive particles is significantly higher at low $\phi$ and in fact {\em decreases} in amplitude with $\phi$ in contrast with noncohesive hard disks as also shown in Fig.~\ref{fig3}(b). It may be further noted that the location of the secondary peaks does not change with $\phi$, whereas those corresponding to the noncohesive disks shift to the left. This is consistent with the fact that it is improbable to form a chain of particles in contact without cohesion. As $\phi$ increases, $g(r)$ appears to grow more similar both in terms of the amplitude of the first peak and the overall form, indicating the increasingly dominant role played by steric interactions in organizing the cluster. 

\begin{figure}

\includegraphics[width=8.6cm]{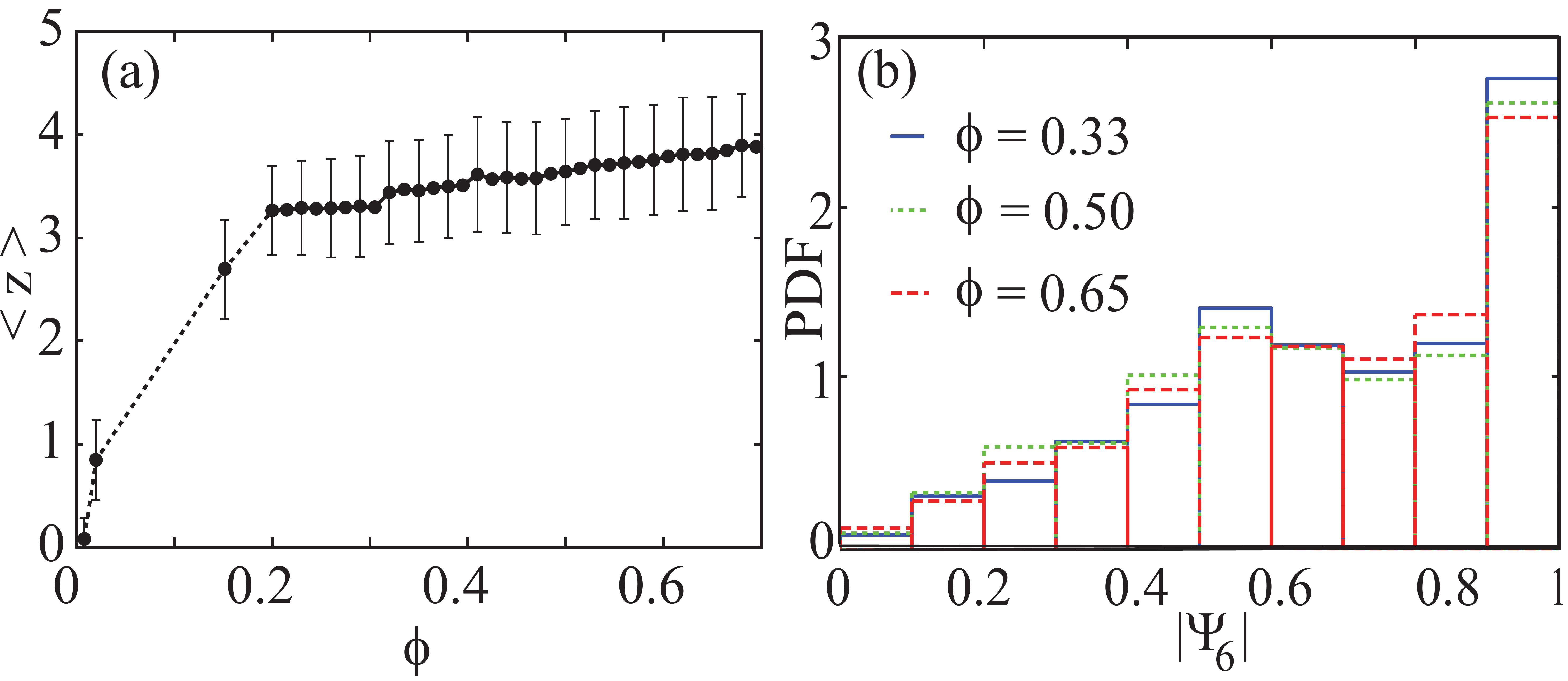}
\caption{\label{fig4} (a) Average number of contacts inside capillary aggregates as a function of $\phi$. Error bars are estimated from statistical dispersion and accuracy of measurement. (b) PDF of the local bond order parameter $\vert\psi_6 \vert $, with a peak near 1 indicating strong hexagonal order.
}
\end{figure}

To understand the connectivity of the aggregates, we obtain the average number of contacts $\langle z \rangle$ between particles (see Fig.~\ref{fig4}(a)). Here, we have used a threshold distance of separation $1.2 d$ for defining a contact between particles to account for cumulative errors in particle tracking and to include all particles which fall in the first peak in $g(r)$. Particles are isolated in the limit $\phi \rightarrow 0$, and therefore  $\langle z \rangle = 0$. It may be noted that $\langle z \rangle$ increases rapidly by $\phi \sim 0.2$ and then does not increase significantly over a broad range of $\phi$. Thus, it appears that once clusters forms, they only grow larger and their local character does not change significantly. The scale over which rapid increase in $\langle z \rangle$ occurs can be estimated based on the following arguments. The average Voronoi cell is an hexagon of area $A_v$ with side given by the average distance between isolated points $l_p$ over $\sqrt{3}$~\cite{Weaire}. For particles randomly distributed inside a circle of radius $R_0$, we get: $\pi {R_0}^2 = N\, \langle A_v \rangle \sim N \dfrac{\sqrt{3}}{2} {l_p}^2$ and ${l_p}^2 \sim \dfrac{\pi\, d^2}{2 \sqrt{3} \phi} $. If ${l_p}<\,4d$, because the experiments are conducted slowly over 600\,s, most of particles are not isolated. Therefore, $ \phi  \gtrsim  {\pi}/{32 \sqrt{3}} \sim 0.06 $, $\langle  z \rangle$ becomes of order one, in qualitative agreement with measurements. Finally, disks with friction should jam for $\left\langle z \right\rangle  \geq 3$ according to the Maxwell criterion in two dimension~\cite{Majmudar}. We observe $\langle z \rangle$ consistent with this value and well below the value of $6$ for a hexagonal crystal. 

A complementary and more subtle method to test the local order, given the hexagonal order indicated by the peaks in $g(r)$,  is using the local bond order parameter $\psi_{6} = \dfrac{1}{N_i} \sum_{j=1}^{N_i} e^{\imath 6 \theta_{ij} } $. Where, $\theta_{ij}$ is the angle between the line joining particles $i$, $j$ and an arbitrary fixed direction, and $N_i$ is the number of nearest neighbors for the particle $i$. Nearest neighbors for each particle are identified as those within $1.2\,d$. We then plot the PDF of $\vert \psi_6 \vert $ in Fig.~\ref{fig4}(b). A peak is observed at $\vert \psi_6 \vert = 1$ showing the prevalence of small hexagonal order. But a broader peak is also observed at $\sim 0.6$ indicating presence of some disorder. Further, the distributions are similar for all the low to high $\phi$ plotted, also confirming that the packing does not evolve significantly with $\phi$ at a local particle level. 

In summary, we investigate the aggregation of cohesive granular matter over a wide range of particle density using a novel experimental system and find unanticipated features in the structures formed. In particular, we find that the observed aggregates at low density show local close packed structure in contrast with cohesionless particles, but the overall heterogeneity is significantly greater than for cohesionless particles, which we characterize with Voronoi cell distributions. Attraction is observed to put the system into a local minima of energy, which introduces a short range order on a scale of few particles. But the structure remains globally disordered, porous and heterogeneous. By applying uniform compression, we show that density is increased by filling of pores. Further we find that the local sub-structure of the particles is on average conserved during the compression. As the density is increased, statistical properties approach that for cohesionless frictional granular as steric interactions increase in importance. Finally, although large scale heterogeneity is reduced at higher densities, global order is never reached and the system buckles far from an ordered state and at densities below jamming threshold for cohesionless particles.

\begin{acknowledgments}

We thank A. Panaitescu, P. Reis, and M. Cloitre for discussions and advice on experiments, and D. Vella for the code to obtain the numerical solution of the Percus-Yevick equation. This work was supported by the National Science Foundation under Grant No. DMR-0605664.

\end{acknowledgments}

\end{document}